\documentclass[12pt]{iopart}

\usepackage{graphicx}
\usepackage{iopams}
\usepackage{subfigure}
\usepackage{overpic}
\usepackage{ulem}
\usepackage{cite}
\usepackage[colorlinks, linkcolor = blue, citecolor = blue, filecolor = blue, urlcolor = blue]{hyperref}

\newcommand{\be}{\begin{equation}}
\newcommand{\ee}{\end{equation}}

\newcommand{\bra}[1]{\ensuremath{\langle #1 |}} 
\newcommand{\ket}[1]{\ensuremath{| #1 \rangle}}

\newcommand{\ovl}[2]{\ensuremath{\langle #1 | #2 \rangle}}

\newcommand*{\eqref}[1]{(\ref{#1})}

\begin{document}
\title[
Stationary quantum coherence and transport in disordered networks]{
Stationary quantum coherence and transport in disordered networks}
\author{Bj\"orn Witt}
\address{Freiburg Institute for Advanced Studies, Albert-Ludwigs University of Freiburg, Albertstr. 19, 79104 Freiburg,
Germany}
\author{Florian Mintert}
\address{Freiburg Institute for Advanced Studies, Albert-Ludwigs University of Freiburg, Albertstr. 19, 79104 Freiburg,
Germany}
\begin{abstract}
We examine the excitation transport across quantum networks that are continuously driven by a constant and incoherent
light source. In particular we investigate the coherence properties of incoherently driven networks by employing recent 
tools from entanglement theory that enable a rigorous interpretation of coherence in the site basis. With these tools at 
hand we identify coherent delocalization of excitations over several sites to be a crucial prerequisite for highly 
efficient transport across networks driven by an incoherent source. These results are set into context with the latest 
discussion of the occurrence and role of coherence in light-harvesting complexes that are exposed to natural incoherent 
sun light.
\end{abstract}
\submitto{\NJP}
\maketitle

\section{Introduction}
Excitation transport across molecular networks is a process that has gained a lot of interest over the last years. Its 
contribution to photosynthesis is crucial as such relies on an efficient transport of excitations across a 
network that wires the antenna complex to a reaction center. The former absorbs the incoming light and creates
excitations which the latter converts into chemical energy. The astonishing efficiency of the transport from 
antenna to reaction center has been known since decades \cite{Chain1977}. Recent two-dimensional spectroscopy 
experiments on biological molecular networks as the Fenna-Matthew-Olson complex (FMO) 
\cite{Engel2007, Panitchayangkoon2010} or the LHC II \cite{Calhoun2009, Schlau-Cohen2009}, however, reflated the 
debate on excitation transport as they revealed signatures of coherent beatings that persist on the timescale of 
several 
hundred femtoseconds to a few nanoseconds. This timescale is comparable to the transport time \cite{Engel2007} which 
raised the question whether genuine quantum coherent effects are one reason for the high efficiency of the transport 
whose underlying mechanism is still a matter of discussion.

An experimental examination in the laboratory is usually based on ultra-short and highly coherent light pulses that are 
applied on the complex and induce the creation of excitations which subsequently propagate across the 
network. Such a scenario is referred to as the transient scenario \cite{Manzano2013, Jesenko2013} as the single 
excitations dynamically propagate from the antenna to the reaction center. Oscillations that can be observed in the 
dynamics \cite{Engel2007, Collini2010} verify the occurrence of coherence whose meaning to the 
transport process is, however, still an open question \cite{Wilde2009, Tiersch2012, Ishizaki2010, Scholes2011}.

Various theoretical approaches to this question suggest coherent delocalization of excitations to be the foundation for 
the high efficiency \cite{Scholak2011c, Zech2012, Ishizaki2009} since it can result in the interference of several 
path-alternatives across the network. If this interference is constructive for the output site, this leads to enhanced 
transport. The maximal extent of this enhancement depends on the maximal number of path-alternatives that interfere 
constructively which, in turn, is limited by the number of sites over which the excitation is coherently delocalized. 
This number therefore characterizes the potential benefit that coherence can have for the transport process.

Recent objections, however, question whether the observations and explanations made for the transient scenario 
based on a coherent excitation process also apply to photosynthesis as it takes place under natural conditions 
\cite{Brumer2012, Mancal2010, Pachon2013}. Instead of an application of a pulsed and coherent light source the 
light harvesting system in vivo is rather exposed to the incoherent and stationary light field of the sun which drives 
the network into a steady state characterized by a 
constant excitation flux from antenna to reaction center. Such a scenario is referred to as the stationary state 
approach \cite{Manzano2013, Jesenko2013} and its relation to the transient scenario as well as the existence and 
role of quantum coherence in the steady states is intensively debated \cite{Fassioli2012, Brumer2012, Mancal2010, 
Jesenko2013, Kassal2013, Pachon2013} as until now no experimental approaches to this question have 
been suggested.

We want to discuss the differences between the transient and the stationary state approach for fully coupled 
random networks by comparing the transport efficiencies of these two scenarios. Furthermore, we examine how coherent 
delocalization of excitations relates to the transport efficiency in an incoherently and continuously driven system in 
order to estimate the role of quantum coherence in the stationary state scenario. To obtain a clear interpretation of 
the extent of coherent delocalization in the mixed stationary network states we employ recent tools from entanglement 
theory \cite{Levi2013} which enable a rigorous characterization of the number of sites over which an excitation is 
coherently delocalized.

\section{Description of the model}
\label{sec:approach}

For the comparison of the transient and the stationary state approach we consider a fully connected network of 
$N$ two-level systems referred to as sites all of which have the same on-site energy. The Hamiltonian for this system 
reads \cite{Adolphs2006}
\begin{equation}
 \label{eq:hamiltonian}
 H = \sum_{i,j=1, \atop i \ne j}^N \frac{\Xi}{|\vec r_i-\vec r_j|^3} \sigma_i^+ \sigma_j^-\ ,
\end{equation}
where $\sigma^+_i$ and $\sigma^-_i$ are the raising and lowering operator on site $i$,
and the interaction decays cubically with the distance between site $i$ and $j$ in accordance with a dipole-dipole 
interaction. We adopt the convention that excitations are fed into the network via site~$1$ and are supposed to 
propagate to site $N$. The two sites define the poles of a sphere, and a random arrangement of the other sites within 
this sphere defines one realization of a network \cite{Scholak2011c}. This model is scale invariant, {\it i.e.} 
increasing the size of the sphere or the interaction constant $\Xi$ can be compensated completely through a proper 
re-scaling of time. We can therefore specify all lengths in terms of multiples of $|\vec r_1-\vec r_N|$ and introduce a 
scaled time $t=\Xi/  |\vec r_1-\vec r_N|^3 \cdot t_r$, where $t_r$ is the actual time, but keep in 
mind that for typical complexes $t=1$ corresponds to $t_r$ of the order of $10^{-13}s$~\cite{Adolphs2006}.

In the stationary state approach the full dynamics including coherent and incoherent contributions is modeled by a
phenomenological master equation. Non-Markovian features can be incorporated in this framework through the use of
time-dependent coupling constants. In the stationary state, however, also these become time-independent 
\cite{Heinz-PeterBreuer2002}, so that the present framework with time-independent rates does not necessarily imply a 
limitation to Markovian
dynamics. More explicitly, the master equation reads
\begin{equation}
\label{eq:equationOfMotion}
\dot{\varrho} = -i \left[ H, \varrho \right]  + \mathcal{L}_{in} \left( \varrho \right) + \mathcal{L}_{rec}\left(
\varrho \right) + \mathcal{L}_{out} \left( \varrho \right) + \mathcal{L}_{deph} \left( \varrho \right)\  .
\end{equation}
The operators $\mathcal{L}_{in}$ and $\mathcal{L}_{out}$ describe the coupling to the external incoherent light field
and to the sink that extracts excitations from the network, respectively. $\mathcal{L}_{deph}$ incorporates the 
dephasing induced by the protein environment and the spin degrees of freedom \cite{Rebentrost2009b, Mohseni2008}, 
whereas $\mathcal{L}_{rec}$ implements recombination, {\it i.e.} the loss of an excitation due to a finite 
lifetime of a site's excited state.

Incoherent feed-in of excitations is modeled by
\begin{equation}
\label{eq:Lin}
  \mathcal{L}_{in} \left( \varrho \right) = \gamma_{in} \left(  \sigma^-_1 \varrho
  \sigma^+_1 - \frac{1}{2} \left\lbrace \sigma^+_1 \sigma^-_1, \varrho \right\rbrace 
  +  \sigma^+_1 \varrho \sigma^-_1 - \frac{1}{2} \left\lbrace
  \sigma^-_1 \sigma^+_1, \varrho \right\rbrace \right) 
\end{equation}
where $\gamma_{in}$ is the rate of absorption from and emission to the incoherent light field. In addition to 
re-emission from the first site into the heat bath all sites can also loose their excitation through recombination as 
induced by
\begin{equation}
\label{eq:Lrec}
  \mathcal{L}_{rec} \left( \varrho \right) = \gamma_{rec}  \sum_{i} \left(  \sigma^-_i \varrho \sigma^+_i -
\frac{1}{2} \left\lbrace \sigma^+_i \sigma^-_i, \varrho \right\rbrace \right)\ ,
\end{equation}
and the coupling to the sink that is described by
\begin{equation}
\label{eq:Lout} 
  \mathcal{L}_{out} \left( \varrho \right) = \gamma_{out} \left( \sigma^-_N \varrho
  \sigma^+_N - \frac{1}{2} \left\lbrace \sigma^+_N \sigma^-_N, \varrho \right\rbrace \right)\ .
\end{equation}
The sink can only withdraw excitation but not feed them back into the network. Additionally to these dissipative 
effects of light field, sink, and recombination the decay of the inter-site coherences is governed by the dephasing 
operator
\begin{equation}
\label{eq:Ldeph} 
  \mathcal{L}_{deph} \left( \varrho \right) = \gamma_{deph}  \sum_{i} \left(  \sigma^-_z \varrho \sigma^+_z -
\frac{1}{2} \left\lbrace \sigma^+_z \sigma^-_z, \varrho \right\rbrace \right)\ ,
\end{equation}
whose prefactor's inverse $\gamma_{deph}^{-1}$ defines an upper limit for the maximal coherence time.

The sink rate $\gamma_{out}$ is chosen to be $\gamma_{out}=20$ as used before in other theoretical studies 
\cite{Rebentrost2009} and in agreement with the typical interaction timescales determined 
experimentally \cite{Owens1987}. In the case of most biological light harvesting complexes it is highly unlikely to 
have more than one excitation in the network at the same time \cite{Fassioli2009} as the inverse of the propagation 
time is usually smaller than the absorption rate. To incorporate the absence of double-excitations in our model we will 
thus choose $\gamma_{in}= 2 \times 10^{-4}$ since we found that for this choice probabilities for double-excitations 
are 
deemed negligible such that we can restrict the following discussion to the zero- and one-excitation subspace.
It should, however, be mentioned that moderate variations of $\gamma_{out}$ or $\gamma_{in}$ do not lead to 
qualitatively different results as long as $\gamma_{in} \ll \gamma_{out}$ holds.

\section{Transport efficiency in the transient and the stationary scenario}
The examination of the transport process requires a quantification of a given network's transport efficiency. Such is 
obtained by introducing two efficiency functions one of which measures the probability for rapid excitation transport 
in the transient case whereas the other one estimates the steady excitation flux to the sink in the stationary state 
scenario. Applying these functions to an ensemble of randomly generated networks enables a comparison of those two 
approaches.

\subsection{Transient and stationary efficiency functions}
In the case of transient dynamics the actual process of extracting the excitation from the network is often not
described explicitly, but efficiency is defined in terms of the probability of the excitation to reach site $N$ 
\cite{Scholak2011c}. A suitable definition is a time-weighted average probability
\be
\label{eq:defTransEff}
E_t=b_t({\cal T}) \int_0^\infty \textrm{d}t \tr\varrho(t)\sigma_N^+\sigma_N^-e^{-\frac{t}{{\cal T}}}\ ,
\ee
where the choice of the time constant ${\cal T}$ permits to gauge the importance attributed to fast transport. The
prefactor $b_t({\cal T}) = {\cal T}^{-1}$ is chosen such that $E_t=1$ is obtained for a hypothetical optimal system that
instantly propagates the excitation to the output without any excitation loss, {\it i.e.\ } $\varrho(t)=\ket{N}\bra{N}$ 
for all times $t$. 

To achieve a meaningful comparison between transient dynamics and steady state properties we need to ensure comparable
time-windows for the excitation to propagate from the input to the output site. Whereas for the transient case this 
time window is defined by $\mathcal{T}$ we can use the recombination rate $\gamma_{rec}$ to provide a limitation on 
the propagation time in the stationary state approach. We will therefore choose the inverse of the recombination 
constant $\gamma_{rec}^{-1}$ in the stationary scenario to approximate the timescale $\cal{T}$ considered in the 
transient case.

Whereas for the transient dynamics efficiency can be defined in terms of the probability for an excitation to
reach site $N$, in the steady state the flux of excitations to the sink is the figure of merit 
\cite{Manzano2013, Manzano2012a}. To identify that sink flux we consider the temporal change of the number of 
excitations in the network by obtaining the expectation value of the number operator $\hat N = \sum_i \sigma_i^+ 
\sigma_i^-$ in the stationary state as
\begin{eqnarray}
\label{eq:statFlux}
\nonumber
\tr \hat N \dot \varrho = -i \underbrace{\tr \hat N \left[ H, \varrho \right]}_{=0}  
+ \underbrace{\tr \hat N \mathcal{L}_{in} \left( \varrho \right)}_{\equiv \tilde \mathcal{J}_{in}} 
+ \underbrace{\tr \hat N \mathcal{L}_{rec}\left( \varrho \right)}_{\equiv - \tilde \mathcal{J}_{rec}}\\ 
+ \underbrace{\tr \hat N \mathcal{L}_{deph}\left( \varrho \right)}_{=0}
+ \underbrace{\tr \hat N \mathcal{L}_{out}\left( \varrho \right)}_{\equiv - \tilde \mathcal{J}_{out}}
\stackrel{!}{=}0 \, .
\end{eqnarray}
One identifies three non-vanishing quantities $\tilde \mathcal{J}_{in}, \tilde \mathcal{J}_{rec}$, and $\tilde
\mathcal{J}_{out}$ which describe the incoming flux, the recombination loss and the sink flux, respectively, and whose
signs have been chosen such that all three quantities are non-negative. With the specific form of ${\cal L}_{in}$
defined above in eq.~\eqref{eq:Lin} the incoming flux can be evaluated to
\be
\label{eq:j_in}
\tilde \mathcal{J}_{in} = \gamma_{in} ( 1 - 2 \tr \varrho \sigma_1^+ \sigma_1^- )\ ,
\ee
while the sink flux can be written as
\be
\tilde \mathcal{J}_{out} = \gamma_{out} \tr \varrho \sigma_N^+\sigma_N^-  \ ,
\label{eq:current}
\ee
where we used the expression for ${\cal L}_{out}$ as specified in eq.~\eqref{eq:Lout}.
As all the three quantities introduced in eq.~\eqref{eq:statFlux} are non-negative, we find the sink flux to be
bounded by the incoming excitation flux which, in turn, is bounded by the injection rate $\gamma_{in}$ according to
eq.~\eqref{eq:j_in}, {\it i.e.} $\tilde \mathcal{J}_{out} \le \tilde \mathcal{J}_{in} \le \gamma_{in}$.
We therefore normalize the sink flux $\tilde {\cal J}_{out}$ with respect to the injection rate, which
yields the stationary transport efficiency
\be
\label{eq:defStatEff}
E_{s} = \frac{\gamma_{out}}{\gamma_{in}} \tr \varrho \sigma_N^+\sigma_N^- \ .
\ee
This quantity will be used in the following for estimating the transport performance in the stationary state.

\subsection{Comparison of the transport efficiency in the transient and the stationary state scenario}
\label{sec:scenarios}
As we have introduced efficiency quantifiers for both of the considered transport scenarios, we can now compare the
efficiency in the stationary state without explicit dephasing according to eq.~\eqref{eq:defStatEff} with the efficiency
in the transient case given by equation eq.~\eqref{eq:defTransEff} for randomly generated networks.
\begin{figure}[ht]
  \centering
  \subfigure[Transport efficiency $E_t$ in a transient scenario as a function of the transport efficiency $E_s$ in a 
  stationary state.]{
    \includegraphics[width=0.46\textwidth]{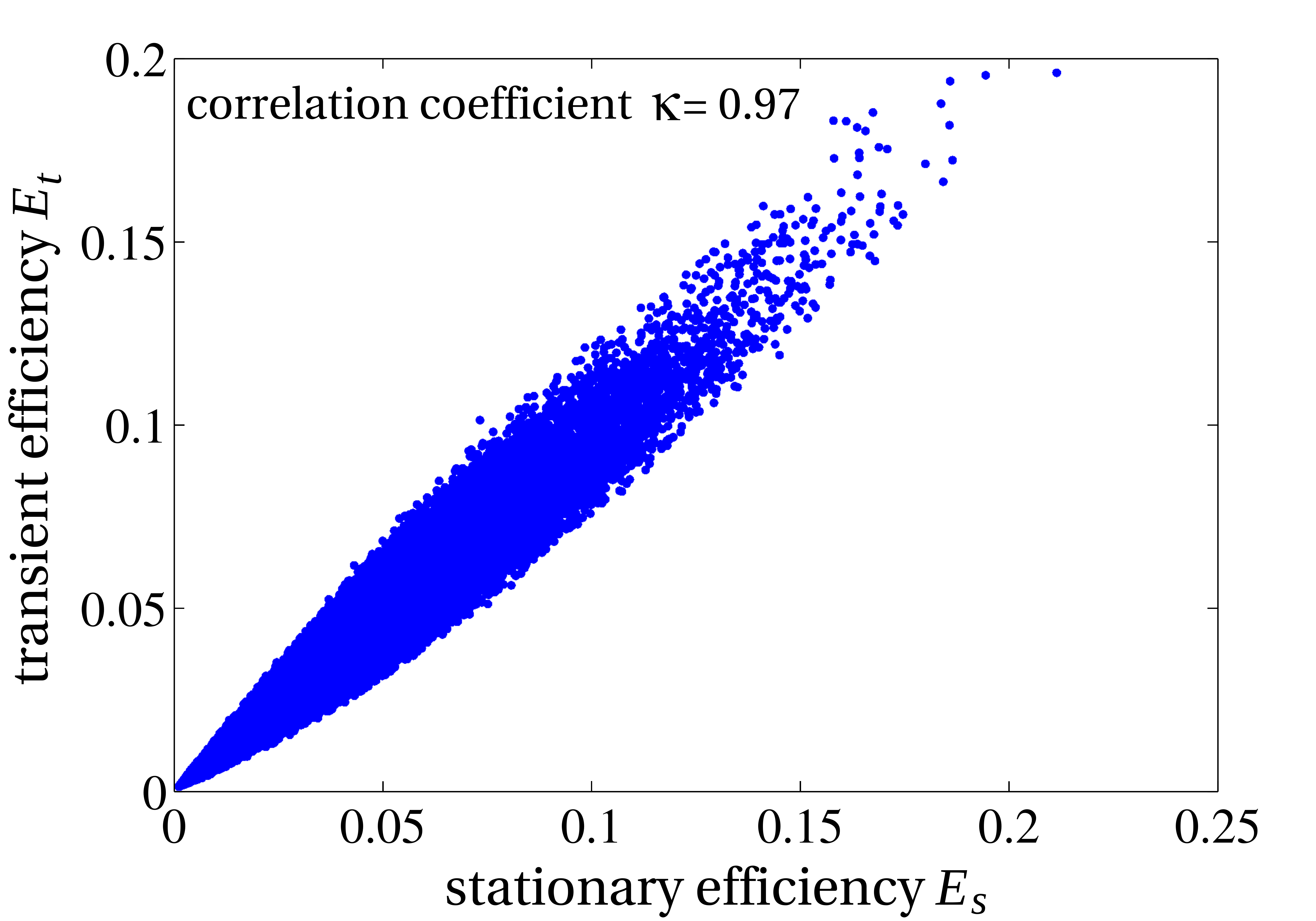}
    \label{fig:comparison}
  }
  \hspace*{0.2cm}
  \subfigure[Correlation coefficient $\kappa$ as a function of the stationary state's excitation lifetime
  $\gamma_{rec}^{-1}$ for various transient lifetimes $\mathcal{T}$.]{
    \includegraphics[width=0.46\textwidth]{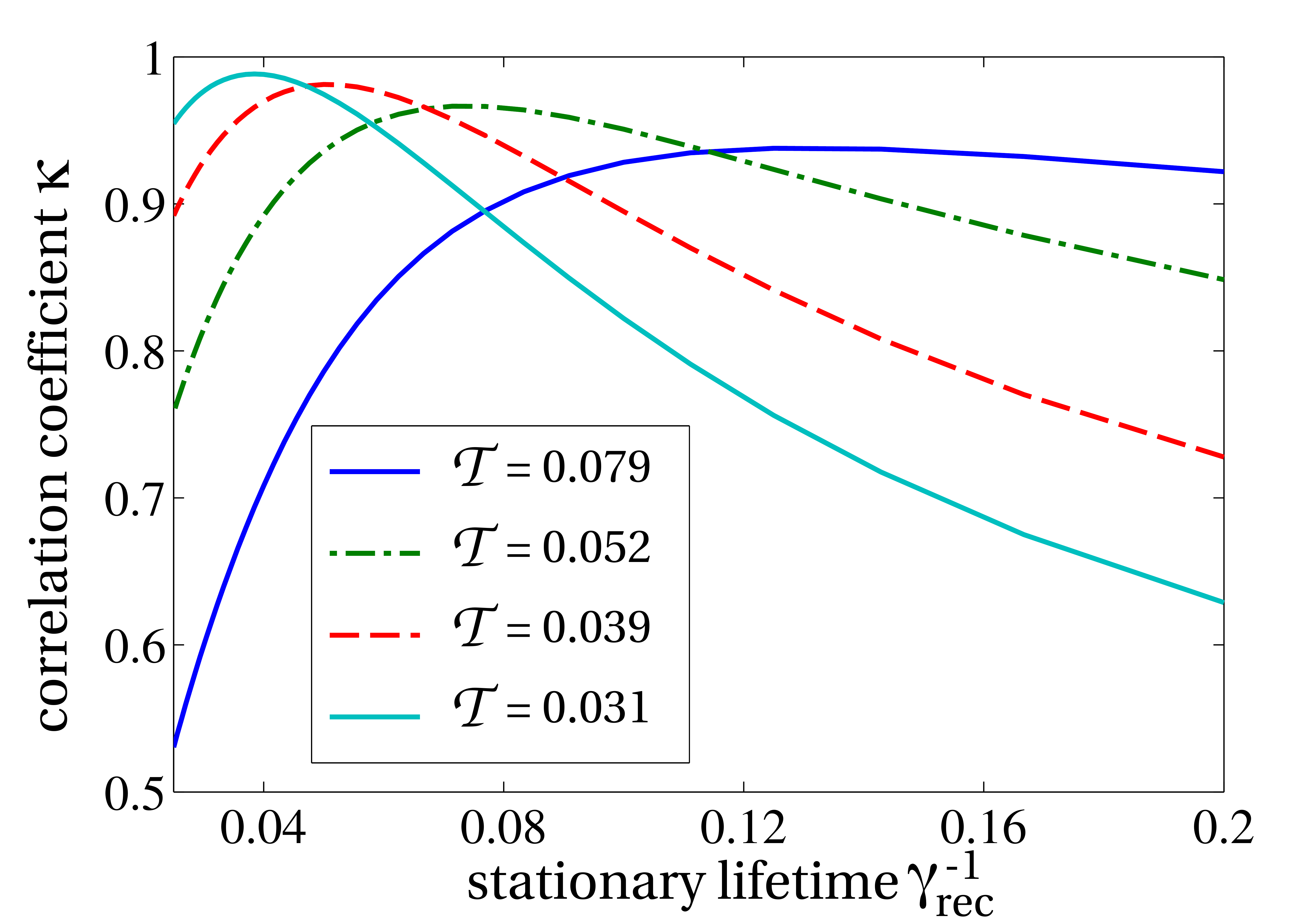}
    \label{fig:correlationCoefficients}
  }
\caption{Correlation of the stationary transport efficiency $E_s$ and its transient analog $E_t$
defined in eq. \eqref{eq:defStatEff} and \eqref{eq:defTransEff} for the same sample of $10^5$ random networks. In the
regime of rapid dynamics, {\it i.e.}\ for excitation lifetimes of $\gamma_{rec}^{-1}=0.05$ and $\mathcal{T}=0.039$,
subfigure {\bf (a)} depicts a strong correlation with a coefficient of $\kappa = 0.97$. This correlation between $E_s$
and $E_t$ decreases for longer lifetimes but a decrease in $\mathcal{T}$ requires a decrease in $\gamma_{rec}^{-1}$ in
order to keep the correlation maximal as depicted in subfigure {\bf (b)} which underlines the compatibility of the two
approaches.}
\end{figure}
Fig.~\ref{fig:comparison} depicts $E_{t}$ as a function of $E_s$ for $10^5$ random systems with $N=7$ sites. A
sufficiently small choice of the transient excitation lifetime ${\cal T} = \frac{1}{40} {\cal T}_{direct} =
\frac{\pi}{80}$ sets the focus on short-time dynamics, {\it i.e.} we only identify those networks as efficient in which
the excitation can reach the output significantly faster than allowed by the direct interaction between input and output
site that takes place on the timescale ${\cal T}_{direct} = \frac{\pi}{2}$. The excitation lifetime in the stationary
state is governed by the inverse of the recombination rate $\gamma_{rec}^{-1} = \frac{1}{20}$. As it can be seen in
fig.~\ref{fig:comparison}, $E_{t}$ and $E_s$ are highly correlated, {\it i.e.} the efficiency in the transient scenario
allows to infer about the efficiency in the stationary case and vice versa. The correlations are not perfect, {\it i.e.}
$E_t$ is not a function of $E_s$ alone, but it can be quantified by the correlation coefficient
\begin{equation}
 \kappa \left( E_t, E_s\right) = \frac{ \left\langle \left(E_t - \left\langle E_t \right\rangle \right)
\left( E_s -
\left\langle E_s \right\rangle \right) \right\rangle }{\sigma_{E_t} \sigma_{E_s}}
\end{equation}
where $\sigma_X=\sqrt{\langle X^2\rangle-\langle X\rangle^2}$ is the standard deviation and $\langle X \rangle$ stands
for the average over the ensemble of random networks. A correlation coefficient of $\kappa=\pm 1$ indicates that the 
value of one quantity determines the value of the other exactly, whereas $\kappa=0$ signifies that the knowledge of one 
does not provide any information about the other. For the data displayed in fig.~\ref{fig:comparison} we obtain a 
correlation of $\kappa \approx 0.97$, {\it i.e.} a close-to-maximal value what substantiates that observations made for 
the transient approach transfer to the stationary scenario essentially
perfectly and vice versa.

One might, however, expect that this compatibility of the two transport scenarios relies on the similarity of 
the transport timescales, {\it i.e.}, it does not apply anymore if $\gamma_{rec} \cdot {\cal T}$ differs substantially 
from unity. To test that, we plot $\kappa$ as a function of the recombination rate's inverse $\gamma_{rec}^{-1}$ for 
different choices of ${\cal T}$ in fig.~\ref{fig:correlationCoefficients}. The case ${\cal T}=\frac{1}{40}{\cal 
T}_{direct}$ corresponding to fig.~\ref{fig:comparison} is depicted in red and a maximum of the correlations for 
$\gamma_{rec} \approx 0.05$ is clearly discernible. The maximum, however, is rather broad, and strong correlations with 
$\kappa>0.9$ are obtained for a wide range of excitation lifetimes in the stationary scenario $0.03 \lesssim 
\gamma^{-1}_{rec} \lesssim 0.1$. That is, the transfer of observations between stationary and transient approach does 
not require precise knowledge of parameters like the recombination rate, but a rough estimate is sufficient for 
qualitative assessments.

A variation of the transient lifetime ${\cal T}$ confirms that optimal correlation is, however, obtained if ${\cal T}$ 
and $\gamma_{rec}^{-1}$ define comparable timescales. Fig.~\ref{fig:correlationCoefficients} depicts that the 
maximum of $\kappa$ is shifted to larger values of $\gamma_{rec}^{-1}$ as ${\cal T}$ is increased what 
clearly underlines the correspondence between these two timescales. The correlation, however, gets less significant for 
longer excitation lifetimes that invoke a consideration of networks with less rapid dynamics. Furthermore, optimal 
correlations are always obtained for $\gamma_{rec} \cdot {\cal T}<1$, {\it i.e.} for the case in which the excitation 
is given a shorter time window in the transient case to reach the output site than in the stationary state scenario.
To appreciate this difference one has to take into account the presence of the sink in the stationary state approach 
which additionally shortens the excitation lifetime. Whereas in the purely coherent case the excitation loss (and 
therefore also the finite excitation lifetime) is only determined by the value of ${\cal T}$, the stationary state's 
excitation lifetime is affected by both recombination and sink drainage. The choice of the sink rate $\gamma_{out}$ does
fundamentally affect the period for which an excitation is able to stay in the network as well as the maximally
obtainable transport efficiency. In the transient approach, however, there is no comparable analog to the sink, what
makes these two concepts differ systematically.

Despite these differences, the correlations between $E_t$ and $E_s$ as well as the correspondence of $\gamma_{rec}^{-1}$
and ${\cal T}$ suggest a major agreement of the transient and the stationary state approach. Efficiency does thus not
primarily depend on the injection mechanism but is a rather universal feature, {\it i.e.}\ a given spacial arrangement
shows a similar transport performance under different feed-in scenarios.

\section{Transport efficiency and quantum coherence}
For the transient scenario it is widely accepted that quantum coherence is a crucial prerequisite for efficient 
transport across molecular networks \cite{Engel2007, Scholak2011c, Zech2012}. Given the agreement of the transient and 
the stationary state scenario in terms of efficiency one might raise the question of whether the similarities go beyond 
the mere transport performance and also hold for the occurrence and role of coherence, \textit{i.e.}\ whether coherence 
in continuously, incoherently driven networks does play the same crucial role for the transport process as assumed for 
the transient case.

The role of coherence for excitation transport is readily illustrated by the double-slit experiment in which a coherent 
superposition of two path-alternatives gives rise to an interference pattern of alternating regions of enhanced and 
reduced arrival probabilities. Increasing the number of coherent path-alternatives through an increasing number of 
slits changes the interference pattern such that it increases the contrast, \textit{i.e.}\ the differences 
in arrival probabilities between spots with constructive and spots with destructive interference. Similar consequences 
also apply to an excitation that can take several path-alternatives in order to propagate from the input to the exit. 
If two of these path-alternatives are in a coherent superposition such that it features constructive interference for 
the output site then this yields a high arrival probability at the output which, in turn, results in an enhancement in 
transport efficiency. In analogy the multi-slit experiment this enhancement can even be higher if there is a 
constructive interference of not only two but three or more path-alternatives. The more different paths are taken 
coherently, the more potential benefit this can have for the transport efficiency.

A coherent superposition of different path-alternatives, however, requires a coherent delocalization of the excitation 
over various sites. The number of sites over which an excitation is coherently delocalized and which we also refer to 
as the extent of coherent delocalization is strongly correlated to number of paths that are in a coherent 
superposition and can thus be expected to also correlate to the optimal transport efficiency.

\subsection{Characterizing coherence in the stationary state}
As discussed before, we can restrict the following discussion to the zero- and one-excitation subspace as due to 
the choice of coupling constants the state amplitudes with more than one excitation are negligible. Since all terms in 
the equations of motion that mediate excitation ({\it i.e.} the Hamiltonian defined in eq.~\eqref{eq:hamiltonian}) 
conserve the number of excitations, the system ground state does not take part in the actual transport so that we only 
need to consider that part of the density matrix that describes a single excitation.

The objective is to characterize to what extent the excitation is coherently delocalized \cite{Smyth2012}.
For that purpose, we project the density matrix onto the single-excitation subspace and apply a renormalization such 
that we obtain
\be
\tilde \varrho=\frac{P\varrho P}{\tr\varrho P}\ \mbox{ with }\ P=\sum_{i=1}^N\ket{i}\bra{i}\ ,
\ee
where $\ket{i}$ denotes the state of the $i^{th}$ site excited and all other sites in the ground state.
Since we expect the benefits and disadvantages of quantum coherence to result from constructive and destructive 
interference of different path-alternatives across the network which, in turn, requires a coherent delocalization of an 
excitation, we will characterize quantum coherence in terms of the number of sites over which an excitation is 
coherently delocalized. An excitation in a pure state is coherently delocalized over $K$ sites if the 
state vector
\be
\ket{\Psi_{KN}}=\sum_i\psi_i\ket{i}\ ,
\ee
contains $K$ finite amplitudes $\psi_i$. Because of the coupling to incoherent reservoirs we, however, always 
face mixed states here for which we have to generalize the concept of $K$-site coherence. The formal generalization is 
fairly straight forward: a mixed state $\tilde \varrho_{KN}$ is considered to feature $K$-site coherence if it can not 
be described as an ensemble of pure states without at least one state-vector with at least $K$-site coherence,
{\it i.e.}
\be
\tilde \varrho_{KN}\neq\sum_{j<K}\sum_ip_{ij}\ket{\Psi^{(i)}_{jN}}\bra{\Psi^{(i)}_{jN}}\ .
\ee
Rigorously identifying $K$-body coherence, on the other hand, is typically rather cumbersome, but given the
formal similarity between $K$-site coherence and $K$-body entanglement in the one-excitation subspace 
\cite{Ishizaki2010}, efficient practical tools can be imported from entanglement theory \cite{Levi2013}. We will employ 
in the following the functions
\begin{eqnarray}
\label{eq:defTau}
\tau_{KN}(\tilde\varrho)=\max_{\{\varphi_i\}}b_{KN}\left(|\bra{\Phi_1}\tilde\varrho\ket{\Phi_2}|-a_{KN}\sum_{i=1}^N\sqrt
{ \bra { \Phi^ {
(i)}_1} \tilde\varrho\ket{\Phi^{(i)}_1}\bra{\Phi^{(i)}_2}\tilde\varrho\ket{\Phi^{(i)}_2}}\right)\ ,\nonumber \\
\end{eqnarray}
with $N$-body product state vectors $\ket{\Phi_1}$, $\ket{\Phi_2}$, $\ket{\Phi^{(i)}_1}$, $\ket{\Phi^{(i)}_2}$
defined in terms of pairs of orthogonal states ({\it i.e.} $\ovl{\varphi_i}{\varphi_i^\perp}=0$) as
\begin{eqnarray}
\begin{array}{rclcrcl}
\ket{\Phi_1}&=&\bigotimes_{i=1}^N\ket{\varphi_i}\ ,&&
\ket{\Phi^{(i)}_1}&=&\bigotimes_{j=1}^{i-1}\ket{\varphi_i}\otimes\ket{\varphi_i^\perp}\bigotimes_{j=i+1}^{N}\ket{
\varphi_i}\ ,\\
\ket{\Phi_2}&=&\bigotimes_{i=1}^N\ket{\varphi_i^\perp},&\mbox{and}&
\ket{\Phi^{(i)}_2}&=&\bigotimes_{j=1}^{i-1}\ket{\varphi_i^\perp}\otimes\ket{\varphi_i}\bigotimes_{j=i+1}^{N}\ket{
\varphi_i^\perp}\ .
\end{array} \nonumber \\
\end{eqnarray}
With the prefactor $a_{KN}$ defined as
\be
a_{KN}=1 / \left( N-K+1\right) \textrm{ for } K \ne 2; \quad a_{KN}=1/N \textrm{ for } K=2
\ee
$\tau_{KN}$ is non-positive for all $N$-body quantum states with a single excitation that do not have at least $K$-site
quantum coherence. The normalization constant $b_{KN}$ is chosen such that $\tau_{KN}$ adopts the value of unity for the
state $\ket{W_{KN}}=\sum_{i=1}^{K}\ket{i}/\sqrt{K}$, {\it i.e.} the state of a system with $N$ sites and an excitation 
that is perfectly coherently delocalized over $K$ sites. Although not strictly necessary for reliable identification of 
$K$-site coherence, we will always perform a numerical optimization over the state-vectors 
$\ket{\varphi_i}$.

\subsection{Coherent excitation transport under incoherent driving}
With these tools at hand, we can now strive for the characterization of coherence properties and their examination with 
respect to the stationary transport efficiencies. That is, we would like to verify if coherent delocalization of the 
excitation is necessary for fast, efficient transport, or whether the identification of this precondition made in the 
transient regime \cite{Scholak2011c} is no longer possible due to the permanent de-cohering impact of the coupling to 
heat baths.

There are two extreme regimes, in which the recombination rate $\gamma_{rec}$ is dominant and negligible respectively
as compared to the inter-site coupling strengths. In the former case where there is no path for the excitation to reach 
the output faster than the limit set by $\gamma_{rec}$ the efficiencies are close to zero. In the latter case where 
$\gamma_{rec}$ is small compared to the average inter-site coupling the excitation will leave through the sink almost 
with certainty as the probability of a backflow to the light field is low and all networks will be characterized 
efficient independently of the question whether transport is coherent or not.

We will therefore focus in the following on the intermediate regime where the value of $\gamma_{rec}$ is of the
same order of magnitude as the typical site-site coupling, {\it i.e.}\ large enough to permit the identification of
fast transport, but still small enough to enable a significant sink flux. We will investigate the coherence properties
of all networks with similar currents, that is we introduce a binning of the $E_s$-axis with windows of width $\Delta
E_s$ centered around $E_s$. We then consider the average $K$-site coherence
\be
\bar\tau_{KN}(E_s)=\langle\tau_{KN}(\tilde\varrho)\rangle_{E_s(\tilde\varrho)\in w_{E_s}}
\label{eq:stat}
\ee
with the average taken over all networks with a current within the window $w_{E_s}=[{E_s}-\frac{\Delta 
E_s}{2},E_s+\frac{\Delta E_s}{2}]$, and the width
\be
\sigma_{KN}(E_s)=\sqrt{\langle(\tau_{KN}(\tilde\varrho)-\bar\tau_{KN})^2\rangle_{E_s(\tilde\varrho)\in w_{E_s}}}
\label{eq:sigma}
\ee
of the distribution of $\tau_{KN}$ within a bin.
\begin{figure}[t]
  \centering 
  \subfigure[Correlation of two-site coherence $\tau_{2,7}$ and stationary transport efficiency.]{
    \includegraphics[width=0.46\textwidth]{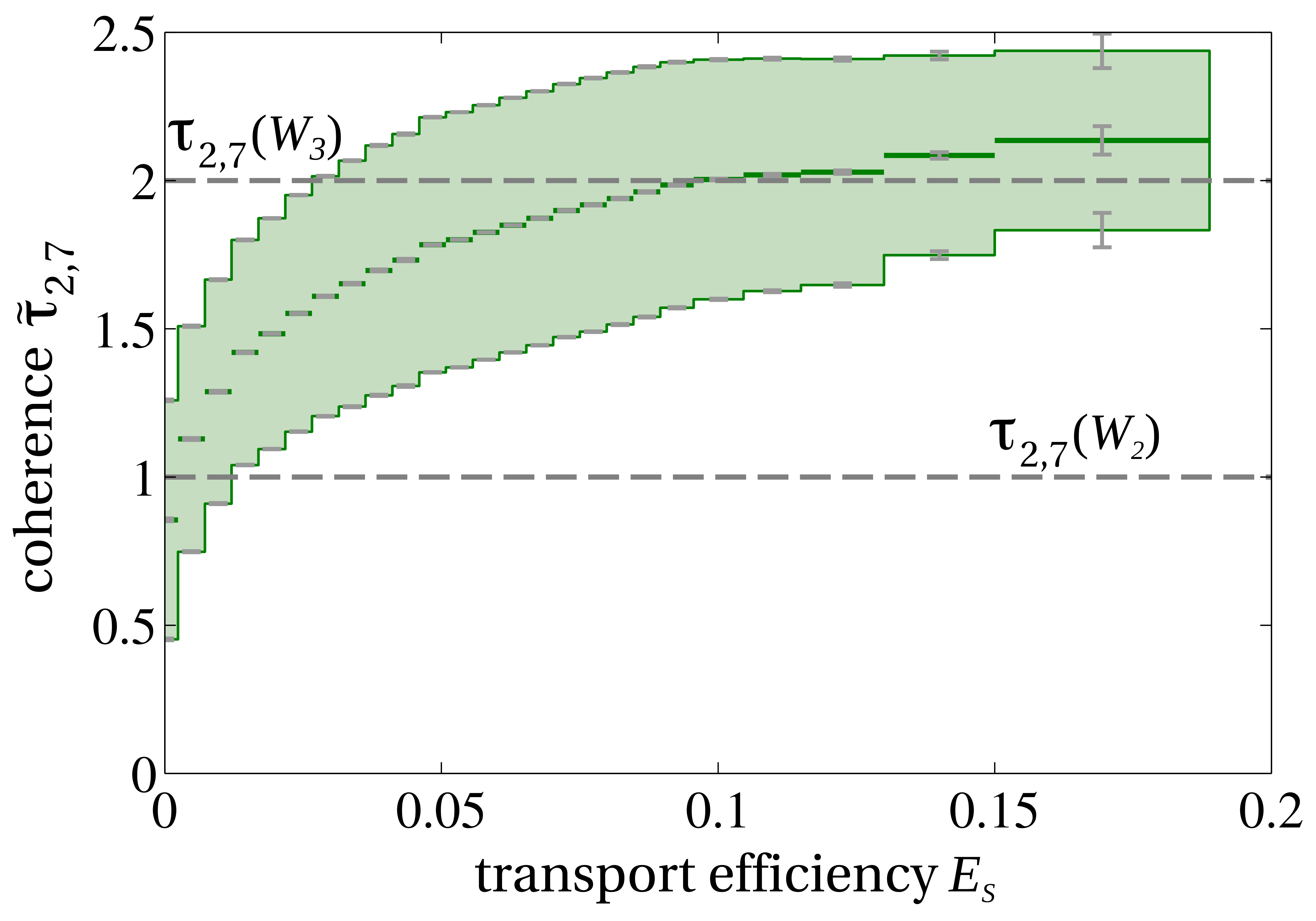}
    \label{img:tau2}     
  }
  \hspace*{0.2cm}
  \subfigure[Correlation of three-site coherence $\tau_{3,7}$ and stationary transport efficiency.]{
    \includegraphics[width=0.46\textwidth]{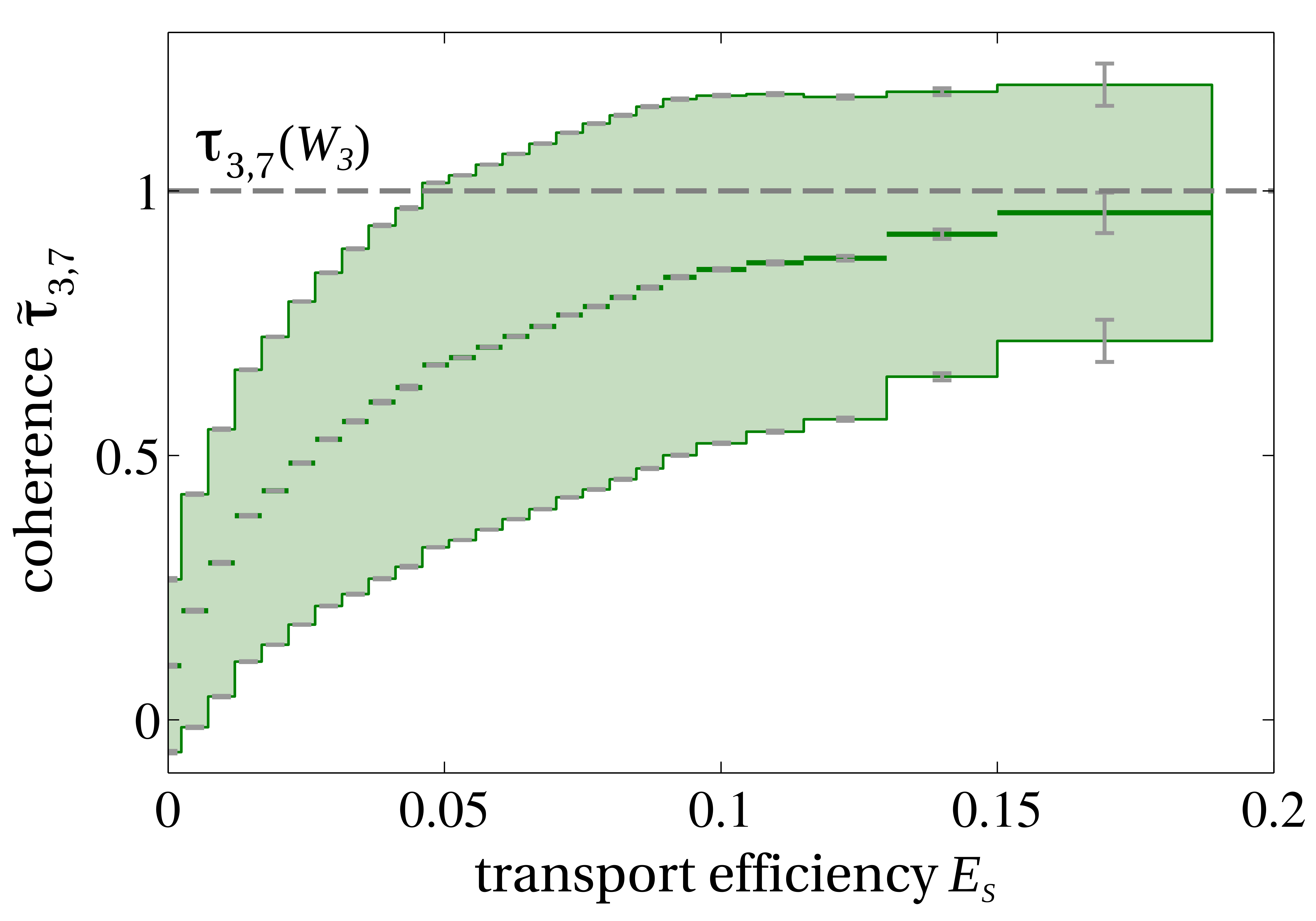}
    \label{img:tau3}     
  }
\caption{Expectation value $\bar \tau_{KN}$ (dark green) and standard deviation $\sigma_{KN}$ (light green) of the 
coherence $\tau_{K,7}$ of $K=2$ (subfigure {\bf (a)}) and $K=3$ (subfigure {\bf (b)}) sites as a function of the 
stationary transport efficiency $E_s$ for randomly arranged dephasing-free networks ($\gamma_{deph}=0$). The error bars 
show the statistical error on the sample mean $\bar \tau_{K,7}$ and on the standard deviation $\sigma_{KN}$. The 
dashed lines represent the coherences $\tau_{K,7}(W_{K'})$ for pure $W_{K'}$-states and indicate the maximal value of 
$\tau_{K,7}$ that can be obtained without $(K'+1)$-site coherent contribution. Whereas rather strong two-site coherence 
can be found for almost all systems the most efficient networks excel by a significant extent of three-site coherence.
}
  \label{img:tau}
\end{figure}
For the absence of dephasing ({\it i.e.}\ for $\gamma_{deph}=0$) fig.~\ref{img:tau2} and \ref{img:tau3} depict the
average two- and three-site coherence defined in eq.~\eqref{eq:stat} for the recombination rate $\gamma_{rec}=20$. 
Since this choice of $\gamma_{rec}$ is of the same order of magnitude as the typical interaction strength between two 
sites, this amounts to rather short-time dynamics. The standard deviation $\sigma_{KN}$ given in eq.~\eqref{eq:sigma} 
is depicted by an bordered area centered around the average value. Additionally to $\sigma_{KN}$ caused by the 
variance of the coherence properties of different random networks in the same efficiency interval we have to 
account for a statistical error $S$ on the sample mean $\bar \tau_{KN}$ as well as on $\sigma_{KN}$ due to the finite 
sample size. Such is indicated by error bars and can be estimated by $S(\bar \tau_{KN}) = \sigma_{KN}/\sqrt{n}$ 
for the average value and $S(\sigma_{KN}) = \Sigma(\sigma_{KN}^2)/(2 \sqrt{\sigma_{KN}})$ for the standard deviation 
where $n$ is the number of networks per bin and 
\be
  \Sigma^2(\sigma_{KN}^2) = \frac{1}{n} \left( \mu_4 - \frac{n-3}{n-1} \sigma_{KN}^4 \right)
\ee
is the variance of the sample variance 
$\sigma_{KN}^2$ \cite{Neter1990} with $\mu_4 = \langle (\bar\tau_{KN} - \tau_{KN}(\tilde \varrho))^4 
\rangle_{E_s(\tilde \varrho) \in \omega_{E_s}}$ being the fourth moment 
about the mean. One finds these statistical errors $S$ to increase for higher efficiencies as efficient networks 
are less likely to be randomly sampled as compared to rather inefficient configurations \cite{Scholak2011c}. Whereas 
sample sizes of roughly $2\cdot10^3$ networks per bin are a sound foundation for obtaining reliable expectation values, 
the bin with the highest efficiency contains less than a hundred networks even though the bin size has been increased.

Despite this statistical error fig.~\ref{img:tau} shows a strong correlation between exciton current and coherence, 
{\it i.e.}\ networks that feature maximal transport also show substantial coherence. Two-site coherence can be detected 
in almost all networks independently of the excitation flux but gains significance as efficiency increases. As the 
value 
of $\tau_{2,7}(W_2)$ obtained for a pure $W_2$-state is significantly exceeded already for fairly low efficiencies 
(see fig. \ref{img:tau2}) one can expect a relevant contribution of three-site coherence, which is confirmed by 
scrutinizing $\bar \tau_{3,7}$ in fig. \ref{img:tau3}. The enhancement of efficiency thus requires a substantial extent 
of coherent delocalization. 

In contrast to the case of $\bar \tau_{2,7}$, however, significant three-site coherence can not be identified in every 
network: In the lowest quarter of the efficiency spectrum, {\it i.e.}\ for $E_s \le 0.05$, values of $\bar\tau_{3,7} \ge 
0.5$ can for example only be found for $46\%$ of all networks. In the most efficient regime $E_s \ge 0.15$, this is the 
case for more than $98\%$ of all systems. This makes three-site coherence a crucial prerequisite for efficient 
transport. 

Coherent delocalization over more than three sites is hardly detectable. As this does not change for larger networks of
$N=9$ or $N=12$, we consider this not to be a finite-size effect due to the limited number of sites in the first 
instance but a consequence of the short time-window provided for establishing coherence which is governed by the 
maximal excitation lifetime $\gamma_{rec}^{-1}$. Positive values for $\bar \tau_{4,7}$
can still be found sporadically but are negligible as compared to $\tau_{4,7}(W_{4})=1$ and the average
value $\bar \tau_{4,7}$ is smaller than the standard deviation for all efficiencies what prevents any statistical
significance. Considering results obtained in a study of the purely coherent transient case 
\cite{Scholak2011c} we must, however, question whether stronger four-site coherence would indeed yield 
more efficient transport. While an examination of the coherent case revealed $K$-site coherence with $K=2$ and $K=3$ to 
be strictly required for high transport efficiency, this relation is weakened drastically for $K \ge 4$. This is 
reasonable as coherent delocalization as a requirement for constructive interference competes with localization on 
the output site in order to obtain optimal efficiency. The transport properties for two- and three-site coherent states 
in the transient scenario do therefore perfectly agree with the results shown here and underline the similarity 
between coherently and incoherently induced transport.

\begin{figure}[t]
  \centering 
  \subfigure[Correlation of two-site coherence $\tau_{2,7}$ and stationary transport efficiency.]{
    \includegraphics[width=0.46\textwidth]{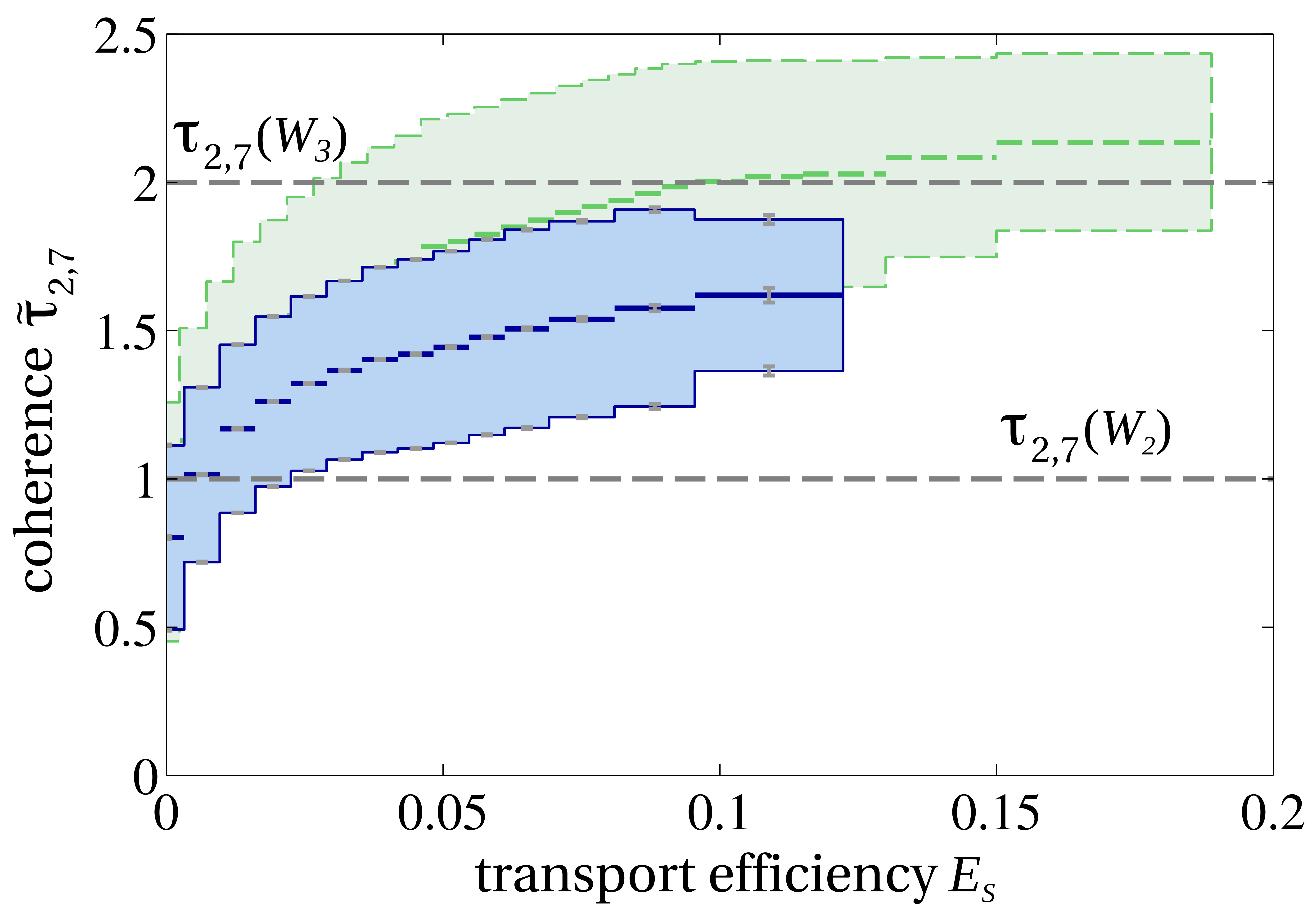}     
    \label{img:tauDeph2}
  }
  \hspace*{0.2cm}
  \subfigure[Correlation of three-site coherence $\tau_{3,7}$ and stationary transport efficiency.]{
    \includegraphics[width=0.46\textwidth]{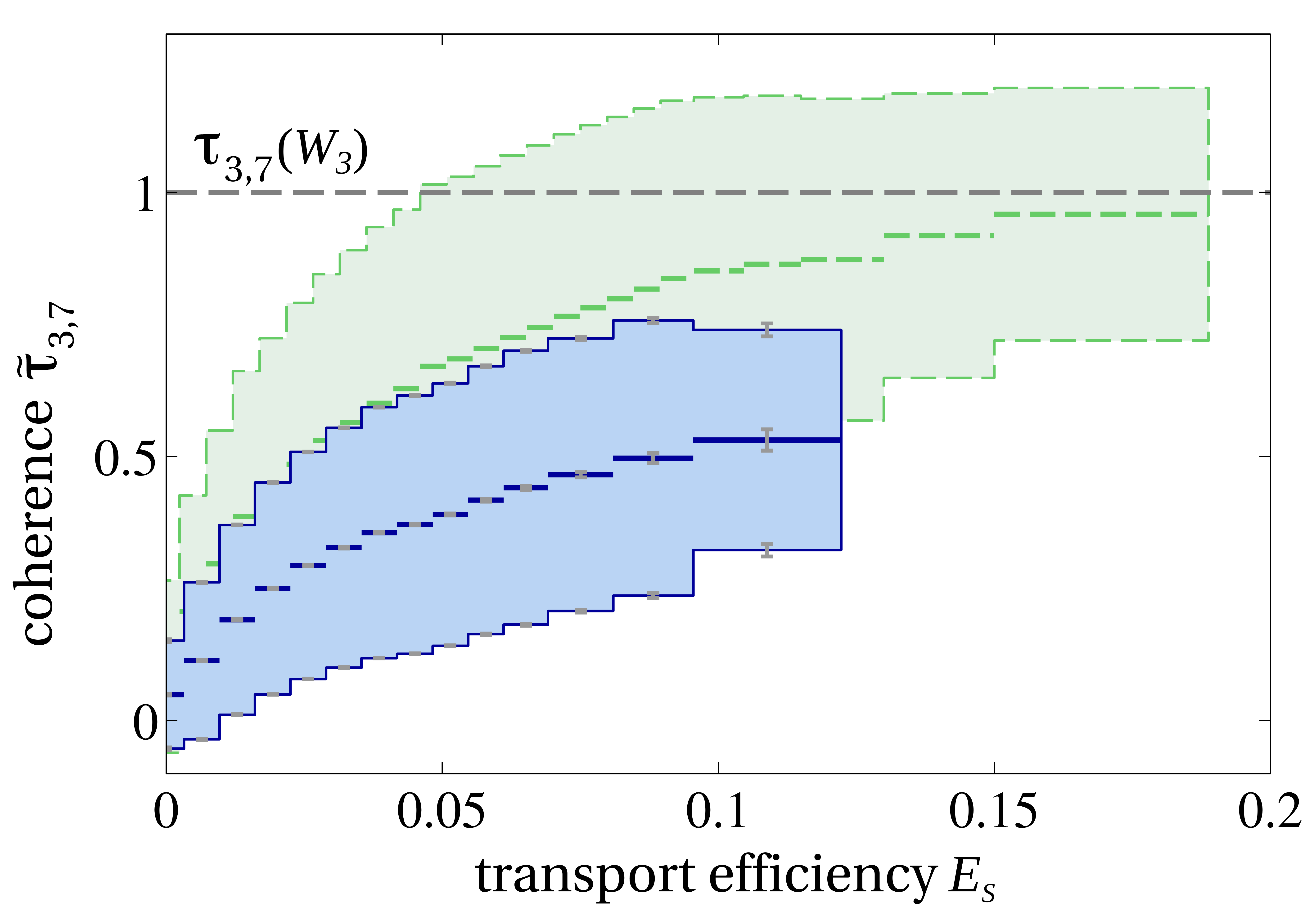}          
    \label{img:tauDeph3}
  }
\caption{Expectation value $\bar \tau_{KN}$ (dark blue) and standard deviation $\sigma_{KN}$ (light blue) of the
coherence $\tau_{K,7}$ of $K=2$ (subfigure {\bf (a)}) and $K=3$ (subfigure {\bf (b)}) sites as a function of the 
stationary transport efficiency $E_s$ for randomly arranged networks exposed to a dephasing of $\gamma_{deph}=10$. The 
error bars show the statistical error on the sample mean $\bar \tau_{K,7}$ and on the standard deviation 
$\sigma_{KN}$. Comparing the results with the dephasing-free scenario depicted in fig. \ref{img:tau} (shown here 
dashed and in green) one finds the correlation between coherence and efficiency to be robust under dephasing. Despite 
the decrease in efficiency of formerly optimal networks the most efficient systems still employ a significant extent of 
coherence which thus can be deemed required for high efficiencies.}
\label{img:tauDeph}
\end{figure}

Despite the dephasing impact of the reservoirs and the recombination, networks of suitable geometry have been 
shown to induce a sufficient amount of coherence which has been identified as a prerequisite for optimal transport. To 
test if the correlation between coherence and efficiency persists under the application of additional dephasing, we 
modify the situation discussed before by changing the dephasing rate to $\gamma_{deph}=10$. The coherence time 
$\gamma_{deph}^{-1}=0.1$ is now comparable to the excitation lifetime $\gamma_{rec}^{-1}=0.05$ that defines the time
window relevant for the system dynamics. This situation thus corresponds to the case of real-world light-harvesting
complexes where the coherence time has been determined experimentally to be of the same order as the transport time
\cite{Engel2007, Panitchayangkoon2010}. A comparison of the transport and coherence properties in the dephasing-free
approach depicted in fig.~\ref{img:tau} and the case of additional dephasing assumed for fig.~\ref{img:tauDeph} suggests
that the incorporation of additional noise does not qualitatively change the relation between coherence and
transport efficiency. Whereas a finite value of $\gamma_{deph}$ leads to reduced coherence the correlation between
$\bar \tau_{K,7}$ and transport efficiency stays rather unaffected. This is in perfect agreement with the decrease
of the maximal efficiency when comparing fig.~\ref{img:tau2} to fig.~\ref{img:tau3}, which shows that optimal networks
loose efficiency when they are exposed to dephasing. Despite the additional noise suitable systems are, however, 
capable to successfully employ interference as long as the system can accumulate a sufficient amount of coherence.

\section{Conclusion}
Based on a comparison of transport efficiencies we found strong similarities between the coherently induced transient 
and the incoherently induced stationary excitation transport suggesting that the underlying transport mechanisms of 
these two scenarios are rather similar. It is this similarity which raises the question if long-lived quantum 
coherence that has been experimentally identified in the transient picture \cite{Engel2007, Panitchayangkoon2010, 
Calhoun2009, Schlau-Cohen2009,  Collini2010} is also relevant for the stationary state approach or whether 
this analogy does not apply due to the incoherent nature of the light source in the latter case \cite{Brumer2012, 
Mancal2010, Pachon2013}.

An application of recent tools from entanglement theory \cite{Levi2013} on incoherently driven random networks 
reveals a strong correlation between coherence and transport efficiency and provides a clear interpretation of the 
considered concept of coherence which in our context always refers to a coherent delocalization of an excitation 
over various pigments. This suggests in particular that suitable molecular networks are capable to exploit
quantum coherent delocalization of excitations for the purpose of efficient excitation transport even if they are 
driven by a thermal light source as the sun. Comparing these outcomes to results obtained for the completely coherent 
and transient transport scenario \cite{Scholak2011c} this permits the conclusion that observations and concepts 
made for the transient scenario can at least qualitatively be transferred to the stationary case and vice versa 
\cite{Jesenko2013}. The coherent beatings detected in two-dimensional spectroscopy experiments can therefore be 
considered as evidence of coherence also in the case of continuous incoherent driving \cite{Fassioli2012, 
Kassal2013, Jesenko2013}.

Our results are robust under the application of additional dephasing, \textit{i.e.}\ under the decrease of coherence 
time. Whereas the latter hinders transport for networks that have originally achieved optimal efficiency, it does not
qualitatively affect the correlation between coherent delocalization and stationary transport efficiency. Consequently, 
networks of suitable geometry are capable to employ quantum coherence in order to obtain optimal transport as long as 
they can induce a sufficient amount of coherence, \textit{i.e.}\ as long as the coherence time is comparable to the 
transport time. That is, if a system's protection against decoherence is good enough to preserve quantum coherence on 
the timescale of the excitation transfer, then this coherence can contribute to an enhanced transport efficiency. 
Based on the observations done in two-dimensional spectroscopy experiments this is the case for various biological 
light-harvesting complexes at physiological temperature \cite{Panitchayangkoon2010} such that our results can be taken 
as a strong argument for nature to employ quantum coherence for transport efficiency enhancement.

\section*{Acknowledgments}
We are grateful to Federico Levi for fruitful discussions and comments. We would like to acknowledge the use of the 
computing resources provided by the Black Forest Grid Initiative and the computing resources provided by bwGRiD 
(\url{http://www.bw-grid.de}), member of the German D-Grid initiative, funded by the Ministry 
for Education and Research 
(Bundesministerium f\"ur Bildung und Forschung) and the Ministry for Science, Research and Arts Baden-Wuerttemberg 
(Ministerium f\"ur Wissenschaft, Forschung und Kunst Baden-W\"urttemberg).

\end{document}